\documentclass[a4paper, preprint, aps, prb, amsmath, amssymb]{revtex4}
\usepackage{hyperref}
\usepackage{amsmath}
\usepackage[utf8]{inputenc}
\usepackage{graphicx}
\usepackage{bm}

\newcommand{\nee}{\ensuremath{n_\mathrm{e}}}
\newcommand{\Te}{\ensuremath{T_\mathrm{e}}}
\newcommand{\Isat}{\ensuremath{I_\mathrm{s}}}
\newcommand{\Isnt}{\ensuremath{I_\mathrm{syn}}}
\newcommand{\neng}{\ensuremath{\overline{n}_\mathrm{e} / n_\mathrm{G}}}

\newcommand{\taud}{\ensuremath{\tau_\mathrm{d}}}
\newcommand{\tauw}{\ensuremath{\tau_\mathrm{w}}}

\newcommand{\avg}[1]{\ensuremath{\langle #1 \rangle}}

\newcommand{\mus}{\ensuremath{\mu \mathrm{s}}}

\newcommand{\Eqnref}[1]{Eq.~\ref{eq:#1}}

\newcommand{\Figref}[1]{Fig.~\ref{fig:#1}}
\newcommand{\Figsref}[1]{Figs.~\ref{fig:#1}}
\newcommand{\figref}[1]{fig.~\ref{fig:#1}}
\newcommand{\Tabref}[1]{Tab.~\ref{tab:#1}}
 
\newcommand{\Secref}[1]{Sec.~\ref{sec:#1}}

\begin{document}

%

\title{Comparison between mirror Langmuir probe and gas puff imaging measurements of intermittent fluctuations in the Alcator C-Mod scrape-off layer}
\author{R. Kube}
\affiliation{Princeton Plasma Physics Laboratory, Princeton University, Princeton, NJ 08543-451, United States of America}
\affiliation{Department of Physics and Technology, UiT The Arctic University of Norway, N-9037 Tromsø, Norway}

\author{A. Theodorsen}
\affiliation{Department of Physics and Technology, UiT The Arctic University of Norway, N-9037 Tromsø, Norway}

\author{O.E. Garcia}
\affiliation{Department of Physics and Technology, UiT The Arctic University of Norway, N-9037 Tromsø, Norway}

\author{D. Brunner}
\affiliation{Commonwealth Fusion Systems, Cambridge, Massachusetts 02139, USA}

\author{B. LaBombard}
\affiliation{MIT Plasma Science and Fusion Center, Cambridge, Massachusetts 02139, USA}
\author{J.L. Terry}
\affiliation{MIT Plasma Science and Fusion Center, Cambridge, Massachusetts 02139, USA}

\begin{abstract} 
Statistical properties of the scrape-off layer (SOL) plasma fluctuations are studied in ohmically heated 
plasmas in the Alcator C-Mod tokamak. For the first time, plasma fluctuations as well as parameters
that describe the fluctuations are compared across measurements from a mirror Langmuir probe (MLP)
and from gas-puff imaging (GPI) that sample the same plasma discharge. This comparison is
complemented by an analysis of line emission time-series data, synthesized from the MLP electron
density and temperature measurements. The fluctuations observed by the MLP and GPI typically display
relative fluctuation amplitudes of order unity together with positively skewed and flattened probability
density functions. Such data time series are well described by an established stochastic framework which model the 
data as a superposition of uncorrelated, two-sided exponential pulses.
The most important parameter of the process is the intermittency parameter, $\gamma = \taud / \tauw$
where $\taud$ denotes the duration time of a single pulse and $\tauw$ gives the
average waiting time between consecutive pulses. Here we show, using a new deconvolution method, 
that these parameters can be consistently estimated  from different statistics of the data.
We also show that the statistical properties of the data sampled by the MLP and GPI diagnostic 
are very similar. Finally, a comparison of the GPI signal to the synthetic line-emission time series suggests that 
the measured emission intensity can not be explained solely by a simplified model which neglects
neutral particle dynamics.
\end{abstract}
\maketitle

\section{Introduction}
The scrape-off layer (SOL) region of magnetically confined plasmas, as used in experiments on fusion
energy, is the interface between the hot fusion plasma and material walls. This region interfaces
the confined fusion plasma and the material walls of the machine vessel. It functions to direct hot
plasma that is exhausted from the closed flux surface volume onto remote targets. In order to
develop predictive modeling capability for the expected particle and heat fluxes on plasma facing
components of the machine vessel, it is important to develop appropriate methods to characterize the
plasma transport processes in the scrape-off layer. 

In the outboard SOL, blob-like plasma filaments transport plasma and heat from the
confined plasma column radially outward toward the main chamber wall. These filaments are
elongated along the magnetic field lines and are spatially localized in the radial-poloidal plane.
They typically present order unity relative fluctuations in the plasma pressure. As they present the
dominant mode of cross-field transport in the scrape-off layer, one needs to understand their collective
effect on the time-averaged plasma profiles and on the fluctuation statistics of the scrape-off
layer plasma in order to develop predictive modeling capabilities for the particle and heat fluxes
impinging on the plasma facing components.

Measuring the SOL plasma pressure at a fixed point in space, the foot-print of a traversing plasma
filament registers as a single pulse. Neglecting the interaction between filaments, a series of
traversing blobs results in a time series that is given by the superposition of pulses. Analysis
of single-point time-series data, measured in several tokamaks, reveals that they feature several
universal statistical properties. First, histograms of single-point time-series  data are well
described by a Gamma distribution 
\citep{graves-2005, horacek-2005, garcia-2013, garcia-2013-jnm, garcia-2015, kube-2016-ppcf, 
theodorsen-2016-ppcf, garcia-2017-nme, garcia-2017-ac, kube-2018, theodorsen-2018-php, kuang-2019}.
Second, conditionally averaged pulse shapes are well described by a two-sided exponential function
\citep{rudakov-2002, boedo-2003, kirnev-2004, garcia-2007-nf, dippolito-2011, banerjee-2012, 
garcia-2013, garcia-2013-jnm, boedo-2014, carralero-2014, kube-2016-ppcf, 
theodorsen-2016-ppcf, garcia-2017-nme, kube-2018}.
Third, waiting times between consecutive pulses are well described by an exponential distribution.
\citep{garcia-2013, garcia-2013, garcia-2013-jnm, garcia-2015, kube-2016-ppcf, garcia-2017-nme,
adamek-2004, kube-2018, theodorsen-2018-php, walkden-2017}.
Fourth, frequency power spectral densities of single point
data time series have a Lorentzian shape. They are flat for low frequencies and decay as a power law
for high frequencies. 
\citep{garcia-2015, garcia-2016, garcia-2017-nme, garcia-2017-ac, theodorsen-2016-ppcf, theodorsen-2017-nf, 
theodorsen-2018-php, kube-2018}
These statistical properties are robust against changes in plasma 
parameters and confinement modes.

These universal statistical properties provide a motivation to model the single-point time-series
data as a super-position of uncorrelated pulses, arriving according to a Poisson process, using a
stochastic model framework. 
\citep{garcia-2012, militello-2016, garcia-2016, theodorsen-2016-php, theodorsen-2017-nf}. 
In this framework, each pulse corresponds to the foot-print of a single plasma filament. Using a
two-sided exponential pulse shape, the stochastic model predicts the fluctuations to be Gamma
distributed. The analytical expression for the frequency power spectral density of this process has
a Lorenzian shape \citep{theodorsen-2017-ps}.
%
The framework furthermore links the average pulse duration time $\taud$ and the average waiting time
between consecutive pulses $\avg{\tauw}$ to the so-called intermittency parameter $\gamma = \taud /
\avg{\tauw}$. This intermittency parameter gives the shape parameter of the gamma distribution that
describes the histogram of data time series and also determines the lowest order statistical moments of
the data time series \citep{garcia-2012}. Recently, it has been shown that using either $\gamma$, or
$\taud$ together with $\avg{\tauw}$, each obtained by a different time series analysis method, allow
for a consistent parameterization of single point data time series \citep{theodorsen-2018-php}. In
order to corroborate the ability of the stochastic model framework to parameterize correctly the
relevant dynamics of single-point time-series data measured in SOL plasmas, and in order to
establish the validity of using different diagnostics to provide the relevant fluctuation
statistics, it is important to compare parameter estimates obtained using a given method and applied
to data sampled by different diagnostics measuring the same plasma discharge.

Langmuir probes and gas-puff imaging diagnostics are routinely used to diagnose scrape-off layer
plasmas. Both diagnostics typically sample the plasma with a few MHz sampling rate and are
therefore suitable to study the relevant transport dynamics. Langmuir probes measure the electric
current and voltage on an electrode immersed into the plasma. The fluctuating plasma parameters are
commonly calculated assuming a constant electron temperature, while in reality the electron
temperature also features intermittent large-amplitude fluctuations, similar to the electron
density \citep{labombard-2007, kube-2018-ppcf, kuang-2019}. 
The rapid biasing that was recently on a scanning probe on Alcator C-Mod 
\citep{labombard-2007, labombard-2014}, the so-called ``mirror'' Langmuir Probe (MLP),
allows measurements the electron density, electron temperature, and the plasma potential on a
sub-microsecond time scale. Moreover, gas-puff imaging (GPI) diagnostics provide two-dimensional images of emission
fluctuations with high time resolution. GPI typically consists of two essential parts. A gas nozzle
puffs a contrast gas into the boundary plasma. The puffed gas atoms are excited by local plasma electrons and
emit characteristic line radiation modulated by fluctuations in the local electron density and
temperature. This emission is sampled by an optical receiver, such as a fast-framing camera or
arrays of avalanche photo diodes (APDs) 
\citep{terry-2001, cziegler-2010, fuchert-2014, zweben-2017-rsi}. 
These receivers are commonly arranged in a two-dimensional field of view and encode the plasma
fluctuations in a time-series of fluctuating emission data. A single channel of the receiver optics
is approximated as data from a single spatial point and can be compared with electric probe 
measurements.

Several comparisons between measurements from GPI and Langmuir probes are found in the literature.
Frequency spectra of the SOL plasma in ASDEX \citep{endler-1995} and Alcator C-Mod 
\citep{zweben-2002, terry-2003} calculated from GPI and Langmuir probe measurements are found to 
agree qualitatively. In other experiments at Alcator C-Mod it was shown that the fluctuations of the 
plasma within the same flux tube, measured at different poloidal positions by GPI and a Langmuir 
probe show a cross-correlation coefficient of more than $60\%$ \cite{grulke-2014}. A comprehensive
overview of GPI diagnostics and comparison to Langmuir probe measurements is given in 
\citep{zweben-2017-rsi}.

\section{Methods}
In this contribution we analyze measurements from the GPI and the MLP diagnostics that were made in
three ohmically heated plasma discharges in Alcator C-Mod, confined in a lower single-null diverted magnetic field
geometry. The GPI was puffing He and imaging the $\mathrm{HeI}\, 587\, \mathrm{nm}$ line in these
discharges. Additionally, we also construct a synthetic signal for the $587\, \mathrm{nm}$ emission
line using the $\nee$ and $\Te$ time-series data reported by the MLP. All plasma discharges had an
an on-axis magnetic field strength of $B_\mathrm{T} = 5.4\, \mathrm{T}$ and a plasma current of
$I_\mathrm{p} = 0.55\, \mathrm{MA}$. The MLPs were connected to the four electrodes of a
Mach probe head, installed on the horizontal scanning probe \citep{brunner-2017-rsi}. In the analyzed
discharges, the scanning probe either performs three scans through the SOL per discharge or dwells 
approximately at the limiter radius for the entire discharge in order to obtain exceptionally long
fluctuation data time series. \Tabref{shot_overview} lists the line-averaged core plasma
density normalized by the Greenwald density \citep{greenwald-2002} and the configuration of the
horizontal scanning probe for the three analyzed discharges. It also lists the average electron 
density and temperature approximately $8$ mm from the last closed flux surface, as measured by the
MLP and mapped to the outboard mid-plane. These values are representative for the SOL plasma.
There is no such data available for discharge 3 since the MLP is dwelled in this case. Since
discharges 2 and 3 feature almost identical plasma parameters, $\avg{ \nee }$ and $\avg{ \Te }$ are
likely to be similar in these two discharges.

\begin{table}
    \begin{center}
    \begin{tabular}{c|c|c|c|c}
        Discharge               & \neng     & $\avg{\nee} / 10^{19}\,\mathrm{m}^{-3}$   & $\avg{\Te} / \mathrm{eV}$ & Probe \\ \hline
        1 (1160616009)          & $0.22$    & $0.19$                                    & $20$                      & scan \\
        2 (1160616016)          & $0.45$    & $0.51$                                    & $15$                      & scan \\
        3 (1160616018)          & $0.45$    & --                                        & --                        & dwell
    \end{tabular}
    \caption{List of the line-averaged core plasma density normalized to the Greenwald density,
             the average electron density and temperature at $\rho \approx 8 \mathrm{mm}$,
             and the operational mode of the horizontal scanning probe.}
    \label{tab:shot_overview}
\end{center}
\end{table}

\begin{figure}
    \centerline{\includegraphics[width=5cm]{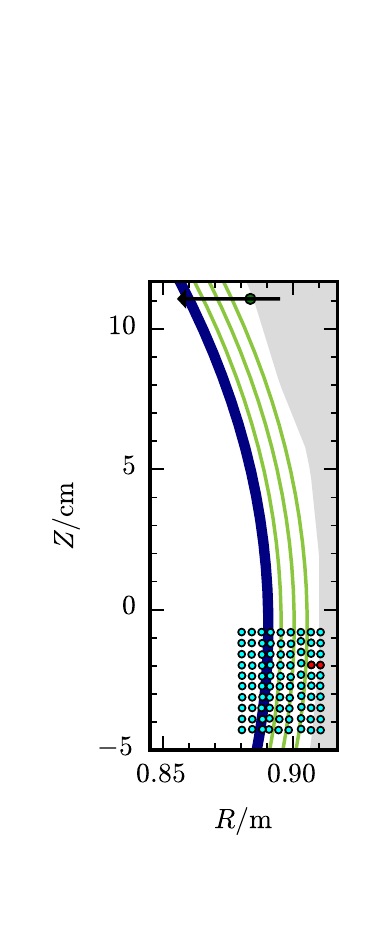}}
    \caption{A poloidal cross-section of Alcator C-Mod's outboard mid-plane section showing the last
             closed flux surface (LCFS) (purple line), magnetic flux surfaces in the SOL (green lines),
             the views of the APDs (cyan dots, red dots show the diode views used in this study), the trajectory
             of the MLP when scanning (black arrow) and the position where the MLP dwells during
             discharge 3 (green dot). }
    \label{fig:cmod_xsection}
\end{figure}

Figure \ref{fig:cmod_xsection} shows a cut-out of the cross-section of the Alcator C-Mod tokamak.
Overlaid are the views of the GPI diodes, the trajectory of the scanning probe head,
as well as the position of the last closed flux surface, obtained from magnetic equilibrium
reconstruction \citep{lao-1985}. The position of the scanning probe in the dwelling position as well as
the position of the GPI views used in this study are highlighted.

\subsection{Calculation of synthetic gas-puff imaging data}
Gas puff imaging diagnostics are routinely used to measure and visualize fluctuations of the
boundary plasma. As realized on Alcator C-Mod, GPI utilizes a vertical stack of 4 ``barrels``,
located approximately $1.5\, \mathrm{cm}$ beyond the outermost column of views, see
\figref{cmod_xsection}, to puff a contrast gas into the boundary plasma.  The line emission arising
from the interaction between the gas atoms and the plasma are captured by a telescope whose optical
axis is approximately toroidal and views the puff with sight lines that are approximately normal to
the $(R,Z)$-plane at the toroidal angle of the nozzle. A fiber optic carries the light imaged by the
telescope to a $9$ by $10$ array of avalanche photo diodes which sample it at $2\, \mathrm{MHz}$
\citep{cziegler-2010}.

The line emission intensity is related to the electron density $\nee$ and temperature $\Te$ as
\begin{align}
    I & = n_0 f(\nee, \Te). \label{eq:GPI_emission}
\end{align}
Here, $n_0$ is the puffed neutral gas density, $\nee$ is the electron density and $\Te$ is the electron
temperature. The function $f$ parameterizes the ratio of the density of particles in the upper level
of the radiative emission to the ground state density times the rate of decay of the upper level. As
discussed in a review by \citet{zweben-2017-rsi}, $f$ has been characterized by a power law
dependence on the electron density and temperature for perturbations around values of 
$\avg{\nee}$ and $\avg{\Te}$ as 
\begin{align}
    f(\nee, \Te) = \nee^{\alpha} \Te^{\beta}. \label{eq:f_parameterization}
\end{align}
The exponents $\alpha$ and $\beta$ also depend on the gas species. Typical values of the fluctuating
plasma parameters in the Alcator C-Mod SOL are given by $5 \times 10^{18}\, \mathrm{m}^{-3} \lesssim
\nee \lesssim 5 \times 10^{19}\, \mathrm{m}^{-3}$ and $10\, \mathrm{eV} \lesssim \Te \lesssim 100\,
\mathrm{eV}$ \citep{labombard-1997, labombard-2001, labombard-2004, kube-2018-ppcf, kube-2019-nme}.

For this parameter range the exponents for HeI are within the range $0.2 \lesssim \alpha \lesssim
0.8$ and $-0.4 \lesssim \beta \lesssim 1.0$. Referring to Figure $7$ in 
\citet{zweben-2017-rsi} we note that in this parameter range $\alpha$ decreases monotonously with
$\nee$ while it varies little with $\Te$ and that $\beta$ decreases monotonically with $\Te$ while
it varies little with $\nee$. Most importantly, $f$ is approximately linear in $\nee$ and $\Te$ for
small $\nee$ and $\Te$ while $f$ becomes less sensitive to $\nee$ and $\Te$ as they increase.

Equation \ref{eq:GPI_emission} relates the measured line emission intensity to the plasma parameters
and is subject to several assumptions. First, the radiative decay rate needs to be faster than
characteristic time scales of the plasma fluctuations, neutral particle transport, and other atomic
physics processes. For the He I $587\, \mathrm{nm}$ line, the radiative decay rate is given by the
Einstein coefficient $A \approx 2 \times 10^{7}\, \mathrm{s}^{-1}$, while the turbulence time scale
is approximately $10\, \mu \mathrm{s}$. 
Second, $n_0$ is assumed to be slowly varying in time so that all fluctuations in $I$ can be ascribed
to fluctuations in $\nee$ and $\Te$.

A synthetic line emission intensity signal is constructed using the emission rate $f$ for the $587\,
\mathrm{nm}$ line of HeI, as calculated in the DEGAS2 code \citep{stotler-1994}, and using the 
$\nee$ and $\Te$ data time-series, as reported by the MLP:
\begin{align}
    \Isnt = f(\nee, \Te). \label{eq:I_syn}
\end{align}
Comparing this expression to \Eqnref{GPI_emission}, we note that the puffed-gas density $n_0$ is
assumed to be constant and absorbed into $\Isnt$. This method for constructing synthetic GPI
emissions is also used in \citet[][]{stotler-2003, halpern-2015-ppcf}.


\subsection{Calculation of profiles}
The fluctuations of the plasma parameters can be characterized by their lower order statistical
moments, that is, the mean, standard deviation, skewness and excess kurtosis. Scanning the Langmuir
probe through the scrape-off layer yields a set of $\Isat$, $\nee$, and $\Te$ samples within a given
radial interval along the scan-path. Here $\Isat$ is the ion saturation current. The center of the
sampled interval is then mapped to the outboard mid-plane and assigned a $\rho_\mathrm{mid}$ value,
corresponding to the distance from the last-closed flux surface. The number of samples within a
given interval depends on the velocity with which the probe moves through the scrape-off layer as
well as the width chosen for the sampling interval. Here, we use only data from the last two probe
scans of discharge $1$ and $2$, as to sample data when the plasma SOL was stable in space and time.


The $\nee$ and $\Te$ data reported by the MLP are partitioned into separate sets for each instance,
where the probe is within $\rho_{\mathrm{mid}} \pm \triangle_\rho$, that is, individually for the
inward and outward part motion and individually for each probe plunge. Thus, for two probe plunges
there are four datasets for $\nee$ and $\Te$ respectively. The lowest order statistical moments are
calculated from the union of these data sets. To estimate the probability distribution function, the
data time series are normalized by subtracting their sample mean and scaling with their respective
root-mean-square value. This procedure was chosen to account for
variations in the SOL plasma on a time scale comparable to the probe reciprocation time scale and 
the delay between consecutive probe plunges. Radial profiles of the lowest order statistical moments
of the GPI data can be calculated using the time series of signals from the individual views.

Skewness $S$ and excess kurtosis $F$ of a data sample are invariant under linear transformations. In
order to remove low-frequency trends in the data time series, for example due to shifts in the
position of the last closed flux surface, $S$ and $F$ are calculated after normalizing the data
samples according to 
\begin{align}
    \widetilde{\Phi} = \frac{\Phi - \avg{\Phi}}{\Phi_\mathrm{rms}}. \label{eq:running_norm}
\end{align}
Here $\avg{\Phi}$ denotes a running average and $\Phi_\mathrm{rms}$ the running root mean square
value. This common normalization allows to compare the statistical properties of the fluctuations
around the mean for different data time series using different diagnostic techniques. In the
remainder of this article, all data time series are normalized according to \Eqnref{running_norm}.

\subsection{Parameter estimation}
It has been shown previously that measurement time series of the scrape-off layer plasma can be modeled
accurately as the super-position of uncorrelated, two-sided exponential pulses.
In the following we discuss how the intermittency parameter $\gamma$, the pulse duration time 
$\taud$, the pulse asymmetry parameter $\lambda$, and the average waiting time between two 
consecutive pulses $\avg{ \tauw }$ are reliably estimated.

The intermittency parameter $\gamma$ is obtained by fitting Equation (A9) in
\citet{theodorsen-2017-ps} on the histogram of the measured time-series data, minimizing the
logarithm of the squared residuals. 
%
%
%
The power spectral density (PSD) for a time series that results from the superposition of 
uncorrelated exponential pulses is given by \citet{garcia-2017-ac},
\begin{align}
    \Omega_{\widetilde{\Phi}}(\omega) & = \frac{2 \taud}
        {\left[1 + \left(1 - \lambda\right)^2 \left(\taud \omega \right)^2 \right] 
         \left[1 + \lambda^2 \left( \omega \taud \right)^2 \right]}. \label{eq:PSD}
\end{align}
Here $\taud$ denotes the pulse duration time and $\lambda$ denotes the pulse asymmetry. The
e-folding time of the pulse rise is then given by $\lambda \taud$ and the e-folding time of the
pulse decay is given by $\left(1 - \lambda\right) \taud$. We note that the PSD of the entire signal
is the same as the PSD of a single pulse. The PSD has a Lorentzian shape, featuring
a flat part for low frequencies and a power-law decay for high frequencies. The point of transition
between these two regions is parameterized by $\taud$ and the width of the transition region is
given by $\lambda$. Note that for very small values of $\lambda$ the power law scaling can be 
further divided into a region where the PSD decays quadratically and into a region where the PSD 
decays as $\left( \taud \omega \right)^{-4}$ \citep{garcia-2017-ac}. For the data at hand, power spectral
densities are calculated using Welch's method. This requires long data time series, which excludes
data from scanning MLP operation.

Data from the MLP are pre-processed by applying a 12-point boxcar window to the data
\citep{labombard-2007}. Assuming that the pulse shapes in the time series of plasma parameters are 
well described by a two-sided exponential function, the MLP registers such pulses as just
this pulse shape filtered with a boxcar window. Since the power spectral density of a superposition
of uncorrelated pulses, i.e. the time series of the plasma parameters, is given by the power
spectral density of an individual pulse \citep{garcia-2017-ac}, the expected power spectrum of MLP
data time series is given by the product of \Eqnref{PSD} and the Fourier transformation of a boxcar
window:
\begin{align}
    \Omega_{\widetilde{\Phi}, \mathrm{MLP}}(\omega) = \Omega_{\widetilde{\Phi}}(\omega) \times \left[ \frac{1}{6 \triangle_{\text{t}} \omega} \sin \left( 6 \triangle_{\text{t}} \omega \right) \right]^2 \label{eq:PSD_MLP}
\end{align}
To estimate the duration time $\taud$ and pulse asymmetry parameter $\lambda$, \Eqnref{PSD} is used
to fit the GPI data and \Eqnref{PSD_MLP} is used to fit the MLP data.

In order to get precise waiting time statistics and the a best estimate of $\tauw$, a method based
on Richardson-Lucy (RL) deconvolution is used \citep{richardson-1972, lucy-1974}. This method was
previously used for a comparison of GPI data from several different confinement modes in Alcator
C-Mod. The method is described in more detail in \citet{theodorsen-2018-php}, here we briefly 
describe the deconvolution. 

By assuming that the dwell MLP and single-diode GPI signals are comprised by a series of uncorrelated
pulses with a common pulse shape $\phi$ and a fixed duration $\taud$, the signals can be written as a
convolution between the pulse shape and a train of delta pulses,
\begin{equation}
    \Phi(t) = \left[\phi * f\right]\left(\frac{t}{\taud}\right),
\end{equation}
where
\begin{equation}
    f(t) = \sum_{k=1}^{K(T)} A_k \delta\left(\frac{t-t_k}{\taud}\right).
\end{equation}
The signal $\Phi$ can be seen as a train of delta pulses arriving according to a Poisson process
$f$, passed through a filter $\phi$. It is therefore called a filtered Poisson process (FPP). For
a prescribed pulse shape $\phi$ and a time series measurement of $\Phi$, the RL-deconvolution can be
used to estimate $f$, that is, the pulse amplitudes $A_\mathrm{k}$ and arrival times $t_\mathrm{k}$.
From the estimated forcing $f$, the waiting time statistics can be extracted. The 
RL-deconvolution is a point-wise iterative procedure which is known to converge to the least-squares
solution \citep{dellacqua-2007}. For measurements with normally distributed measurement noise, the 
$n+1$'th iteration is given by \citep{witherspoon-1986, pruksch-1998, dellacqua-2007, tai-2011}
\begin{equation}
    f^{(n+1)}(t) = f^{(n)}(t) \frac{\left[\Phi * \widehat{\phi}\right](t)}{\left[f^{(n)} * \phi * \widehat{\phi}\right](t)},
\end{equation}
where $\widehat{\phi}(t) = \phi(-t)$. For non-negative $\Phi$ and $f^{(0)}$, each following
iteration will be non-negative as well. The initial choice $f^{(0)}$ is otherwise unimportant, and
has here been set at constant unity. For consistency with PSD estimates of $\taud$ and $\lambda$
(see \Secref{results}), we use a two-sided exponential pulse function with $\taud = 20 \,\mu \text{s}$
and $\lambda = 1/10$ for the GPI data, and a two-sided exponential pulse function with 
$\taud=10\, \mu\text{s}$ and $\lambda = 1/25$ convolved with the 12-point window for the
MLP data. The deconvolution procedure is robust to small deviations in the pulse shape.

The deconvolution algorithm was run for $10^5$ iterations, after which the L$^2$-difference between the
measured time series and the reconstructed time series  was considered
sufficiently small. The result of the deconvolution resembles a series of sharply localized,
Gaussian pulses, so a peak-finding algorithm is employed in order to extract pulse arrival times and
amplitudes from the deconvolved signal. The window size of the peak finding algorithm is chosen to
give the best fit to the expected number of events in the time series, resulting in window sizes of
$7.5\, \mu \mathrm{s}$ (\Isat), $0.9\, \mu \text{s}$ (\nee), $6.3 \, \mu \text{s}$ (\Te), 
$4.5\, \mu \text{s}$ (GPI, for the view at $90.7\, \text{cm}$) 
and $7.5\, \mu \text{s}$ (GPI, for the view at $91 \, \text{cm}$). 
The deconvolution procedure finds $85001$, $200332$, $101815$, $30574$ and $17343$ pulses in these
time series, respectively.

In order to test the fidelity of the process, a synthetic time series consisting of a pure FPP 
has been subjected to the deconvolution procedure as well. This time
series has the same sampling time, $\taud$ and $\lambda$ as the GPI time series, with $\gamma=2$.
In this case, a window of $5.5\,\mu \text{s}$ gives the best fit to
the expected number of events and the procedure finds 48011 events (the true number of events in the
synthetic time series is 50000).

\begin{figure}
    \centerline{\includegraphics{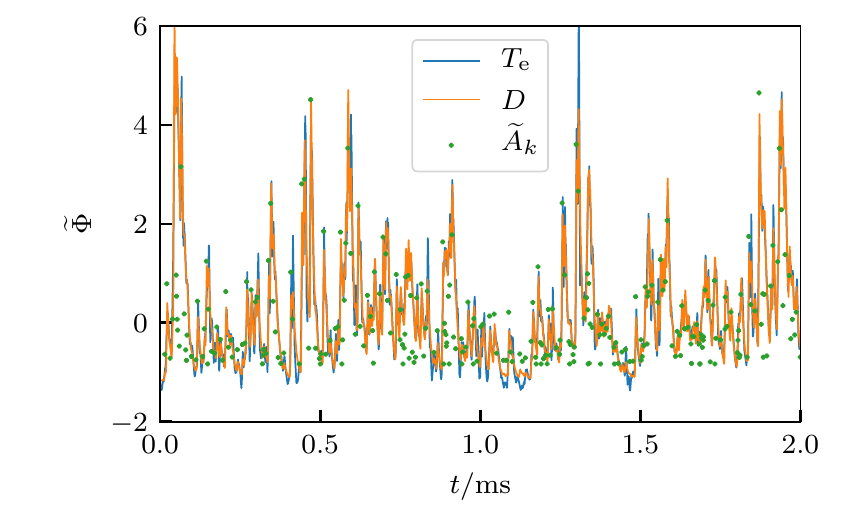}}
    \caption{Excerpt of original (as measured by the MLP) and reconstructed {\Te} signals. The blue
             curve gives the original signal \Te, the green dots indicate arrival times $t_\mathrm{k}$
             and normalized amplitudes for the pulses $A_\mathrm{k}$ and the orange curve gives the
             reconstructed signal $D$. All signals are normalized such as to have zero mean and unit
            standard deviation.}
    \label{fig:deconv-ts-Te}
\end{figure}

\begin{figure}
    \centerline{\includegraphics{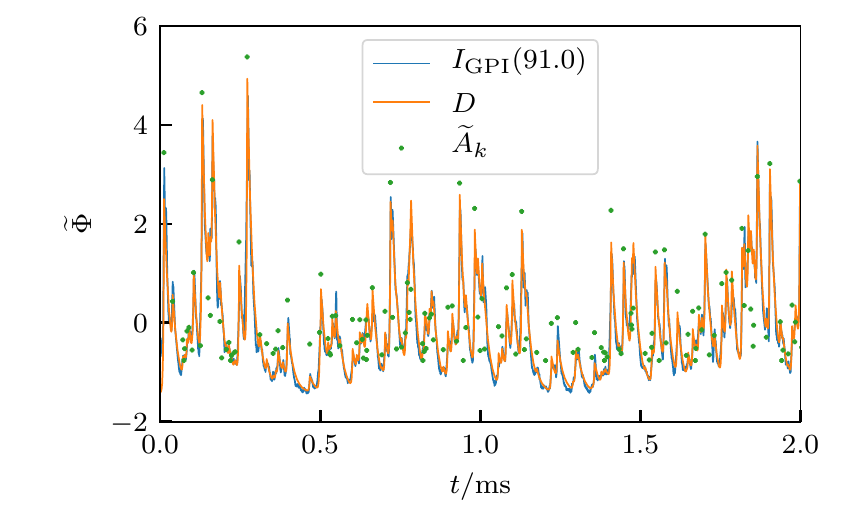}}
    \caption{Excerpt of original and reconstructed GPI signals at $R=91\,\text{cm}$. The blue curve
             gives the original signal $I_\mathrm{GPI}$, the green dots indicate arrival times 
             $t_\mathrm{k}$ and normalized amplitudes $A_\mathrm{k}$ for the pulses and the orange
             curve gives the reconstructed signal $D$. All signals are normalized such as to have
             zero mean and unit standard deviation.}
    \label{fig:deconv-ts-G91}
\end{figure}

Example excerpts of the reconstructed time series are presented in \Figsref{deconv-ts-Te} and
\ref{fig:deconv-ts-G91}. In both figures, the blue lines give the original time series, normalized
according to \Eqnref{running_norm}. The green dots indicate the pulse arrival times and amplitudes which
are the output of the deconvolution procedure described above. The amplitudes have been normalized
by their own mean value and standard deviation. By convolving the estimated train of delta pulses
with the pulse shape, the full time series is reconstructed. The result of this reconstruction
is given by the orange lines. Overall, the reconstruction is excellent.

\section{Results}\label{sec:results}

\subsection{Statistical properties of synthetic GPI intensity}
\begin{figure}
    \centerline{\includegraphics{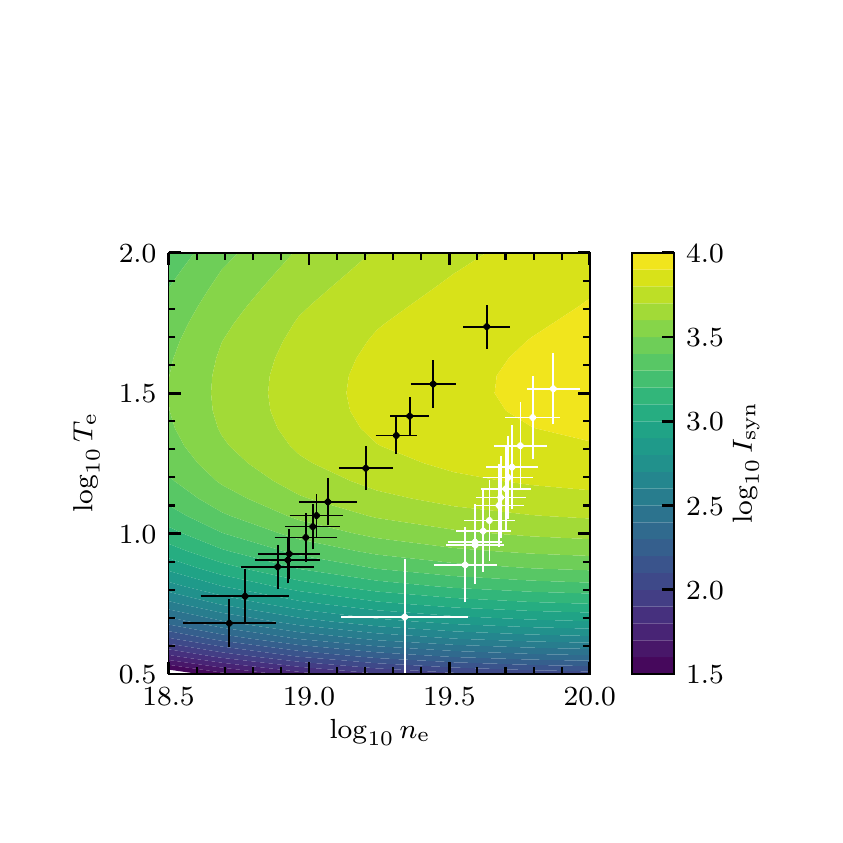}}
    \caption{Synthetic emission rate for the HeI $587\, \mathrm{nm}$ line as a function of the electron  
             density and temperature. Overplotted are the average $\nee$ and $\Te$ values reported
             by the MLP at different $\rho$ positions in discharge 1 (black markers) and 2 (white markers).
             The error bars are given by the respective root-mean-square values.}
    \label{fig:emission_rates} 
\end{figure}

Synthetic GPI emission rates are calculated according to \Eqnref{I_syn}, using data reported from
the MLP in discharges 1 and 2. Figure \ref{fig:emission_rates} color codes the emission intensity
given by \Eqnref{f_parameterization} with data reported by the MLP overlaid. Discharge 1 features a
scrape-off layer that is colder and less dense than the SOL plasma in discharge 2. Furthermore, the
gradient scale-lengths of the $\avg{ \nee }$ and $\avg{ \Te }$ profiles are shorter in discharge 1
\citep{kube-2019-nme}. Thus, the range of reported $\nee$ and $\Te$ values in discharge 1 (black
markers) is larger than the range reported in discharge 2 (white markers). The contour lines suggest
that both $\partial \Isnt / \partial \Te$ and $\partial \Isnt / \partial \nee$ are larger over the
parameter range relevant for discharge 1 than they are for discharge 2. Consequently, variations in
the amplitude of the plasma parameters $\nee$ and $\Te$ are mapped in a non-linear way to variations
in the amplitude of $\Isnt$ and the local fluctuation exponents $\alpha$ and $\beta$ can not be
used. Appendix A gives a more detailed discussion regarding the local exponent approximation.

\begin{figure}
    \centering
    \begin{minipage}{0.45\textwidth}
        \includegraphics[width=\textwidth]{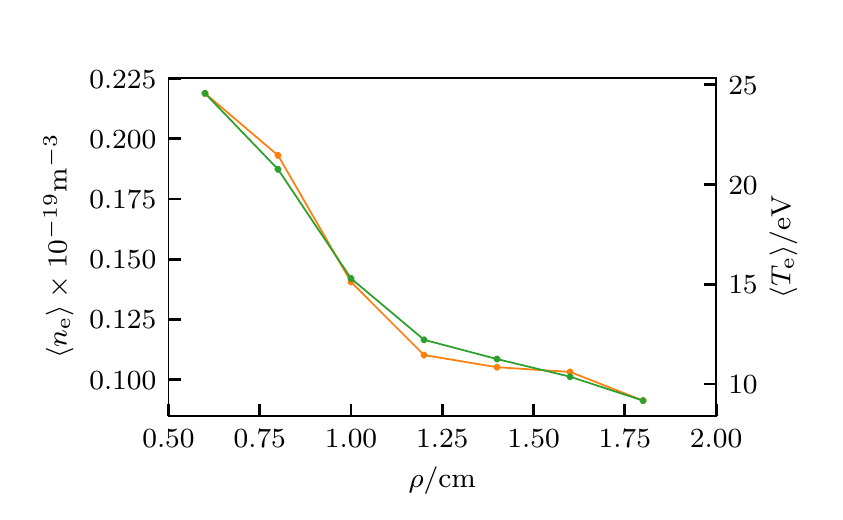}
        \caption{Average}
        \label{fig:profiles_shot1_avg}
    \end{minipage}
    \hfill
    \begin{minipage}{0.45\textwidth}
        \includegraphics[width=\textwidth]{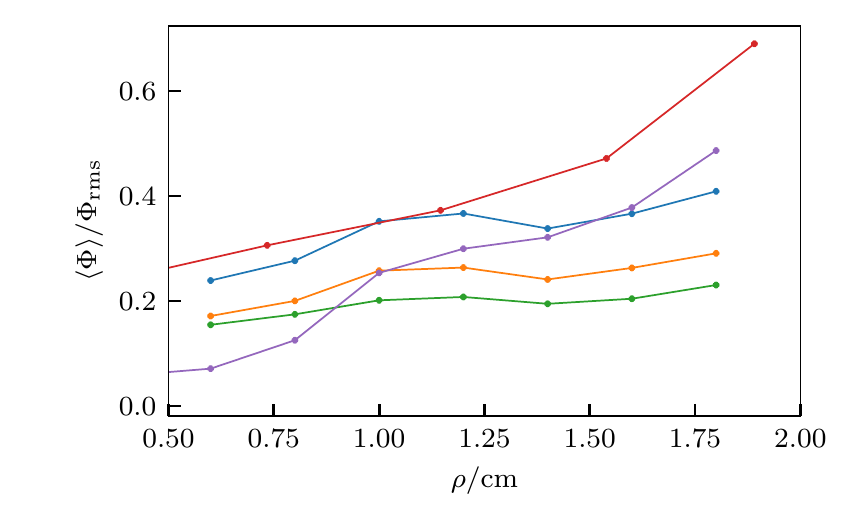}
        \caption{Relative fluctuation level}
        \label{fig:profiles_shot1_rfl}
    \end{minipage}
    \newline
    \begin{minipage}{0.45\textwidth}
        \includegraphics[width=\textwidth]{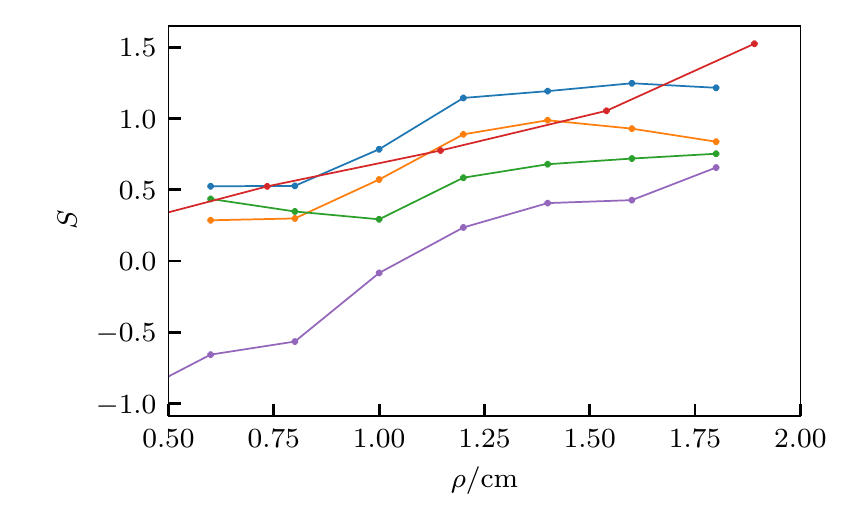}
        \caption{Skewness}
        \label{fig:profiles_shot1_skw}       
    \end{minipage}
    \hfill
    \begin{minipage}{0.45\textwidth}
        \includegraphics[width=\textwidth]{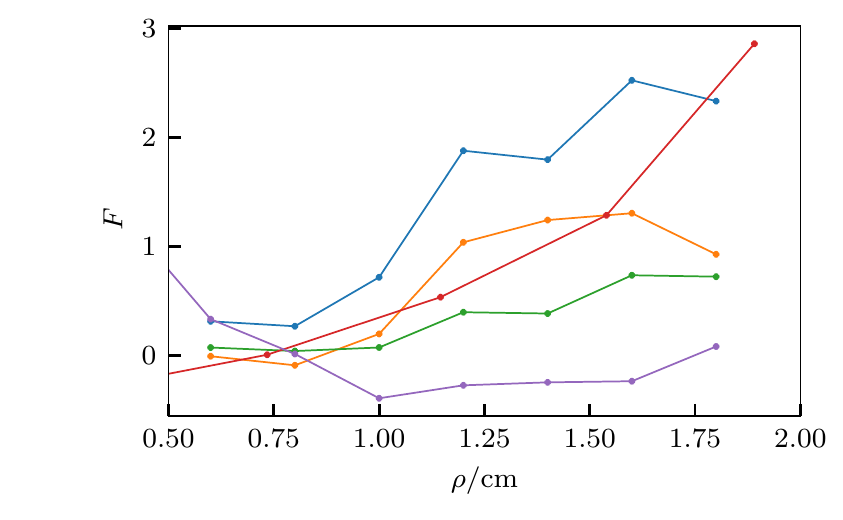}
        \caption{Relative fluctuation level}
        \label{fig:profiles_shot1_krt}
    \end{minipage}
    \newline
    \begin{minipage}{0.45\textwidth}
        \includegraphics[width=\textwidth]{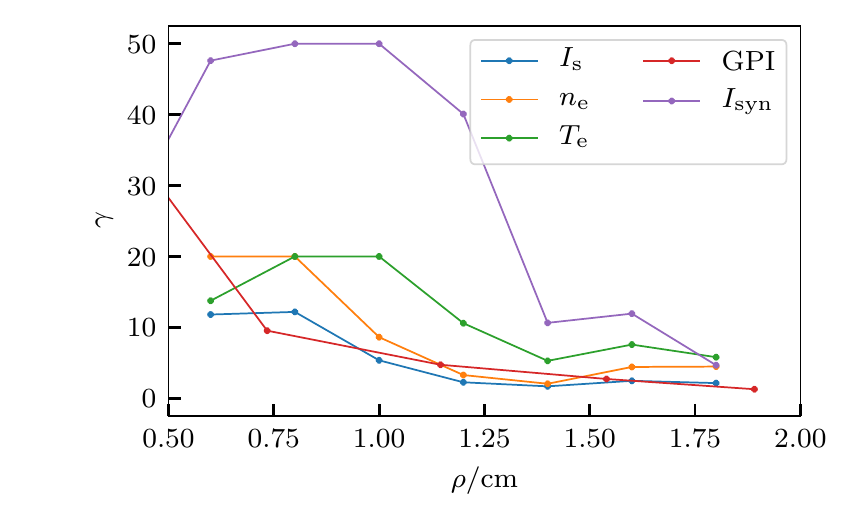}
        \caption{Intermittency parameter}
        \label{fig:profiles_shot1_gmm}
    \end{minipage}  
    \caption{Radial profiles of various quantities sampled in discharge 1. The color
             legend in subplot \ref{fig:profiles_shot1_gmm} applies to all subplots.}
    \label{fig:profiles_shot1}
\end{figure}

We now compare the lowest-order statistical moments of the different signals. Figure
\ref{fig:profiles_shot1} shows radial profiles of the mean, the relative fluctuation level, skewness and
intermittency parameter for the relevant MLP data ($\nee$, $\Te$, and $\Isat$), the GPI data, as
well as the synthetic GPI data ($\Isnt$). Looking at the profile of the average values of $\nee$ and
$\Te$, shown in \ref{fig:profiles_shot1_avg} we note that the scale lengths of both quantities is 
almost identical. Both $\nee$ and $\Te$ decay sharply for $\rho \lesssim 1\, \mathrm{cm}$. With 
larger distance from the LCFS their profiles feature a larger scale length. 
Both MLP and GPI data feature a fluctuation level of up to
$0.5$ times their respective mean. This relative fluctuation level increases with distance from the
LCFS. The relative fluctuation level of the $\Isnt$ data also increases with
$\rho$ but is less than the fluctuation level of the GPI data (by factors of $\sim 0.85$ and 
$\sim 0.3$) over the profile.
Coefficients of sample skewness for the MLP and the GPI data are positive, comparable in magnitude
and increase with $\rho$. The synthetic data features negative sample skewness for $\rho \lesssim
1\, \mathrm{cm}$ but are positive and increasing for $\rho \gtrsim 1\, \mathrm{cm}$. For both, MLP
and GPI data, $F$ increases from approximately $0$ at $\rho \approx 0.5\, \mathrm{cm}$ to larger
positive values for $\rho \approx 1.5\mathrm{cm}$. $F$ calculated using $\Isnt$ data is approximately
zero over the entire range of $\rho$. The lowest panel of \Figref{profiles_shot1} shows the
intermittency parameter $\gamma$, obtained by a fit on the histogram of data sampled in a given
$\rho$ bin. Both, MLP and GPI data feature a large value of $\gamma \gtrsim 10$ for
$\rho \lesssim 1 \mathrm{cm}$. This implies that the PDFs closely follow a
normal distribution, which is consistent with small values of $S$ and $F$. For larger $\rho$ values
the data features positively skewed and flattened  histograms, a feature captured by the smaller
$\gamma$ value and compatible with the larger estimates of $S$ and $F$. For the synthetic data,
$\gamma$ is estimated to be larger than $10$ over the entire range of $\rho$. This implies that
these samples closely follow a normal distribution, which is compatible with nearly vanishing 
skewness and excess kurtosis of this data.


\begin{figure}
    \centerline{\includegraphics{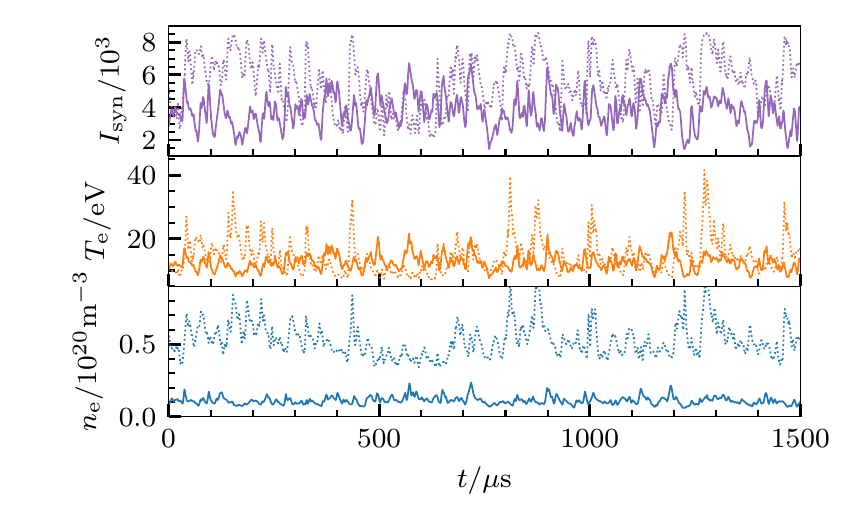}}
    \caption{Time series of $\nee$ (lowest panel), $\Te$ (middle panel),
    and synthetic GPI data (top panel) for discharge 1 (full line) and 2 (dashed line). Data are taken
    in the first interval where the probe scans from $\rho = 1.3$ to $1.2\, \mathrm{cm}$.}
    \label{fig:timeseries_shot12}
\end{figure}

\begin{figure}
    \centerline{\includegraphics{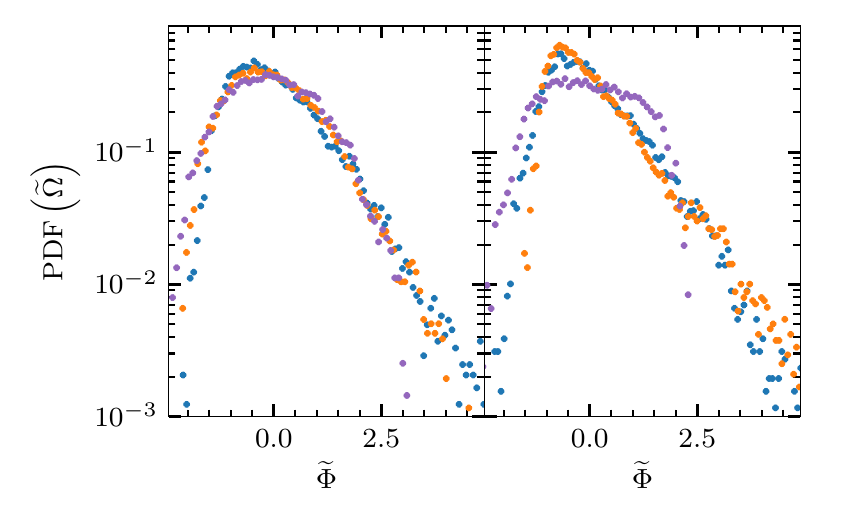}}
    \caption{Histogram of $\widetilde{n}_\mathrm{e}$ (blue dots), 
             $\widetilde{T}_\mathrm{e}$ (orange dots),
             and $\widetilde{I}_\mathrm{syn}$ (purple dots) 
             for discharge 1 (left) and 2 (right). Data are taken
             in all intervals where the probe scans from $\rho = 1.2$ to $1.3\, \mathrm{cm}$.}
    \label{fig:histogram_shot12}
\end{figure}

While the radial profiles of the lowest order statistical moments calculated using MLP and GPI data
agree qualitatively the profiles of the $\Isnt$ data show large discrepancies. The relative
fluctuation level of the $\Isnt$ data is comparable to the relative fluctuation level of the
$\Isat$, $\nee$, $\Te$ and the GPI data, while $S$, $F$, and $\gamma$ calculated using $\Isnt$ data
correspond to a near-gaussian process. \Figref{timeseries_shot12} shows $\nee$, $\Te$ and $\Isnt$
time series. The waveforms of the $\nee$ and $\Te$ data present intermittent and asymmetric
large-amplitude bursts for both discharge 1 and 2. Peaks in the $\Isnt$ on the other hand appear
with a somewhat smaller amplitude relative to the quiet time between bursts and with a more
symmetric shape. Histograms of the corresponding data, shown in \Figref{histogram_shot12},
corroborate this interpretation. For the data sampled in discharge 1 (full lines in
\Figref{timeseries_shot12} and the left panel in \Figref{histogram_shot12}), histograms of the
$\nee$ and $\Te$ data are asymmetric with elevated tails for large-amplitude events. The histogram
of the $\Isnt$ data on the other hand features no elevated tail for large amplitude events. For
$\widetilde{I}_\mathrm{syn} \gtrsim 2.5$ the histogram is approximately zero. For discharge 2
(dashed lines in \Figref{timeseries_shot12} and the right panel in \Figref{histogram_shot12}),
the histogram of the $\Isnt$ data appears symmetric and features a plateau around
$\widetilde{I}_\mathrm{syn} = 0$ without a pronounced peak.

The different fluctuation statistics can be understood by referring to \Figref{emission_rates}. For
one, $\Isnt$ is more sensitive to $\Te$ fluctuations than to $\nee$ fluctuations, that is, $\partial
\Isnt / \partial \Te > \partial \Isnt / \partial \nee$ within relevant ranges of $\nee$ and $\Te$.
Furthermore, both scaling exponents $\alpha$ and $\beta$ may vary significantly over the range of a
single large-amplitude burst, as indicated by the error bars. Since $\nee$ and $\Te$ fluctuations
are strongly correlated and feature similar pulse shapes \citep{kube-2018-ppcf}, \Eqnref{I_syn} does
not result in a perfectly scaled pulse shape of the input signals. For example, when assuming a
two-sided exponential pulse for $\nee$ and $\Te$ as input for \Eqnref{I_syn}, the resulting pulse
shape is not a two-sided exponential pulse, but rather a boxcar-like pulse as the saturation levels
of $\nee$ and $\Te$ are reached.

%

\subsection{Statistical properties of MLP and GPI data}
\begin{figure}
    \centerline{\includegraphics{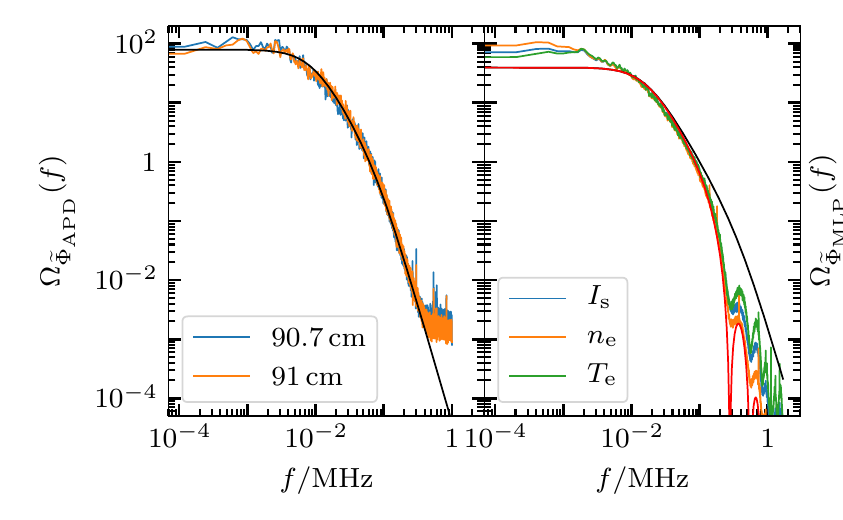}}
    \caption{The left panel shows the PSD of the GPI data from discharge 3 and \Eqnref{PSD} using
             parameters from a least-squares fit. The right panel shows PSDs of the MLP signals and 
             both, \Eqnref{PSD_MLP} (red line) and \Eqnref{PSD} (black line) evaluated using 
             parameters found from a least-squares fit.}
    \label{fig:psd_fit}
\end{figure}

Figure \ref{fig:psd_fit} shows the frequency power spectral densities calculated from MLP and GPI
data sampled in discharge 3. The PSDs of the GPI data from the different radial positions, shown in
the left panel, are almost identical. They are flat for low frequencies,
$f \lesssim 5\, \mathrm{kHz}$, before transitioning into a broken power law
decay for high frequencies. A least squares fit of \Eqnref{PSD} on the data (black line) yields 
$\taud = 20\, \mu \mathrm{s}$ and $\lambda \approx 0.1$ and describes the PSDs of the signals
perfectly over more than 4 decades. 
 
PSDs of the MLP data ($\Isat$, $\nee$, and $\Te$) appear similar in shape to the PSD of the GPI
data, except that for high frequencies, $f \gtrsim 0.2\, \mathrm{MHz}$, a ``ringing'' effect can be
observed. This is due to internal data processing of the MLP, which smooths data with a 12-point
uniform filter as discussed above \citep{kube-2018-ppcf}. Fitting \Eqnref{PSD_MLP} on the data yields $\taud = 10\, \mu
\mathrm{s}$ and $\lambda = 0.04$. The red and black line in the right-hand panel show \Eqnref{PSD_MLP}
and \Eqnref{PSD} respectively with these parameters. While \Eqnref{PSD_MLP} describes the
Lorentzian-like decay of the experimental data as well as the ''ringing'' effect at high
frequencies, it underestimates the low frequency part of the spectrum, 
$f \lesssim 10^{-2}\, \mathrm{MHz}$. This is addressed by the deconvolution procedure.

Summarizing the parameters found by fitting the GPI and MLP data, we find $\taud = 20\, \mu
\mathrm{s}$ and $\lambda = 1/10$ for GPI data and $\taud = 10\, \mu \mathrm{s}$ and $\lambda = 1/25$
for MLP data. In other words, the MLP observes shorter pulses that are more asymmetric than the GPI.
Since the GPI measures light emissions from a finite volume (that is at least the 4 mm diameter
spot-size times the toroidal extent of the gas cloud) and pulses in the signal are due to radially-
or poloidally- 
propagating blob structures, it can be expected that the registered pulses in the signal appear more
smeared out, compared to those from the Langmuir probes, which measure plasma parameters at the probe
tips. No such ``pulse smearing'' pollutes the MLP signals. This may be the reason for the
difference found for the $\taud$ and $\lambda$ parameters.

\begin{figure}
    \centerline{\includegraphics{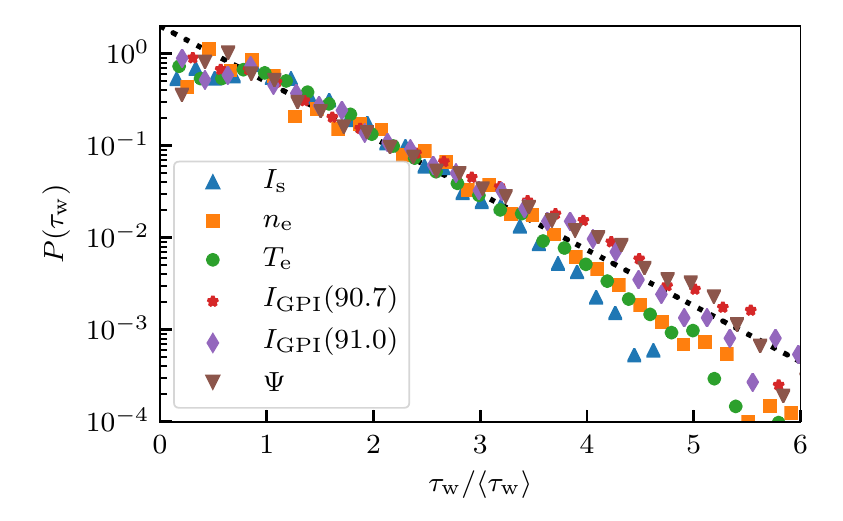}}
    \caption{Probability density functions of waiting times obtained from deconvolving the GPI and
             MLP time series. The synthetically generated time series is indicated by $\Psi$. The
             black dotted line indicates exponential decay.}
    \label{fig:deconv-compare-tw-pdf}
\end{figure}

\begin{figure}
    \centerline{\includegraphics{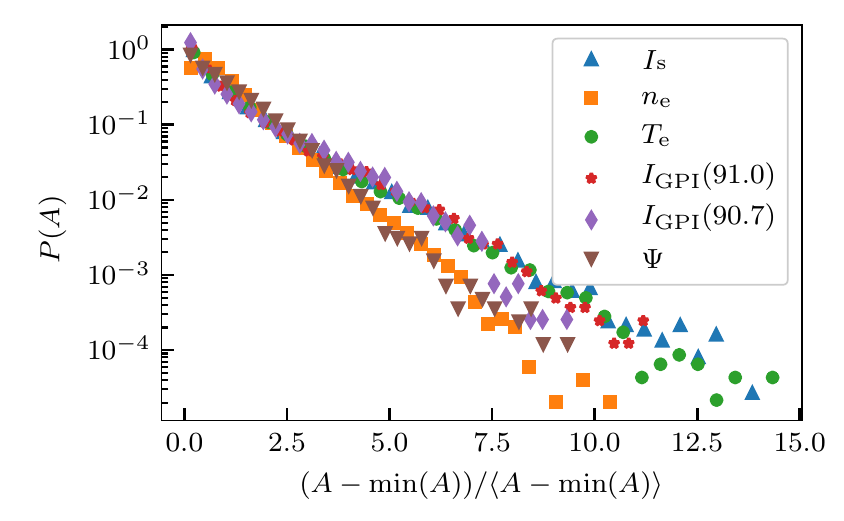}}
    \caption{Probability density functions of the pulse amplitudes obtained from deconvolving the
             GPI and MLP time series. The synthetically generated time series is indicated by $\Psi$.}
    \label{fig:deconv-compare-A-pdf}
\end{figure}

\begin{figure}
    \centerline{\includegraphics{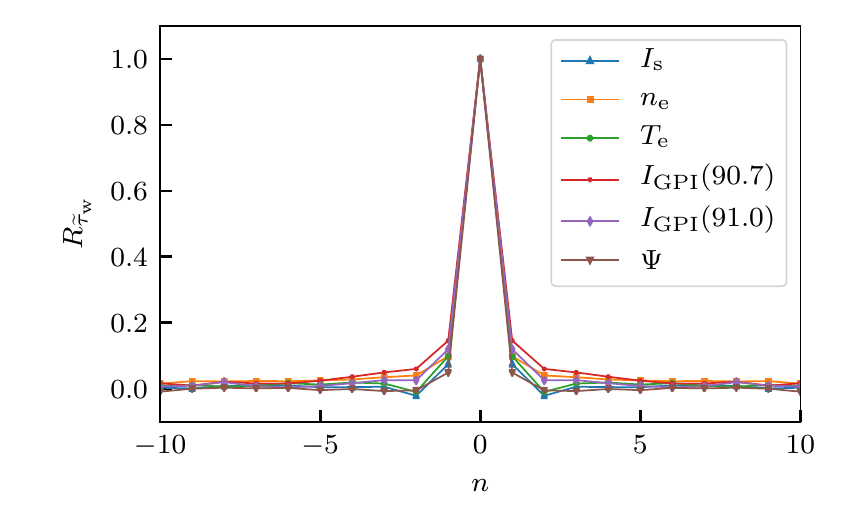}}
    \caption{Auto-correlation functions of the waiting times obtained from the deconvolution 
             procedure. Here, $n$ gives the number of waiting times away from the reference waiting
             time. The synthetically generated time series is indicated by $\Psi$.}
    \label{fig:deconv-compare-tw-corr}
\end{figure}

\Figsref{deconv-compare-tw-pdf} - \Figref{deconv-compare-tw-corr} show the results of the
deconvolution procedure, starting with the PDF of the waiting times. The brown triangles give the
estimated waiting times of the synthetically generated signal, while the black dotted line indicates
an exponential decay. The GPI waiting time distribution conforms very well to the exponential decay of
the synthetic time series for the entire distribution. The MLP waiting time distributions decay
exponentially over at least two decades in probability. All waiting time distributions have lower
probability of small waiting times ($\tauw / \avg{ \tauw } \lesssim 0.8$) compared to an exponential
distribution, an artifact of the non-zero $\taud$ and the peak finding algorithm. This is also true 
for the synthetic time series.

\Figref{deconv-compare-A-pdf} shows the PDF of the pulse amplitudes obtained by applying the 
deconvolution procedure. The pulse amplitudes are approximately exponentially distributed for all
analyzed signals. The $\nee$ data and the synthetic GPI data both appear sub-exponential. On the other
hand, the distribution of pulse amplitudes both GPI data time series appears to be identical to the
distribution reconstructed from the $\Isat$ and $\Te$ data. For small and large amplitudes, the 
plotted PDFs show deviations from an exponential function. The deviation for large amplitudes is due
to the finite size of the data time series. Deviations for small amplitudes are also observed in 
other measurement data \citep{theodorsen-2018-php}.

In \Figref{deconv-compare-tw-corr}, the autocorrelation function of the consecutive waiting times is
presented.  Here, $R_{\widetilde{\tau}_\text{w}}[n] = R_{\widetilde{\tau}_\text{w}}[k, k+n] =
\langle \widetilde{\tau}_{\text{w},k} \, \widetilde{\tau}_{\text{w},k+n} \rangle$, and
$\tau_\text{w}$ is normalized by subtracting its mean value and dividing by its standard deviation.
This function is very close to a delta function, indicating that consecutive pulses are
uncorrelated and thus supporting the assumptions of pulses arriving according to a Poisson process.

Together, these results indicate that the waiting times derived from the GPI and MLP data follow the
same distribution and are consistent with exponentially distributed and independent waiting times.
This further justifies using the stochastic model framework. The estimated average waiting times are
presented in \Tabref{parameter_est}, and give $\gamma$-values consistent with those obtained from
fits to the histograms of the time series.

\begin{figure}
    \centerline{\includegraphics{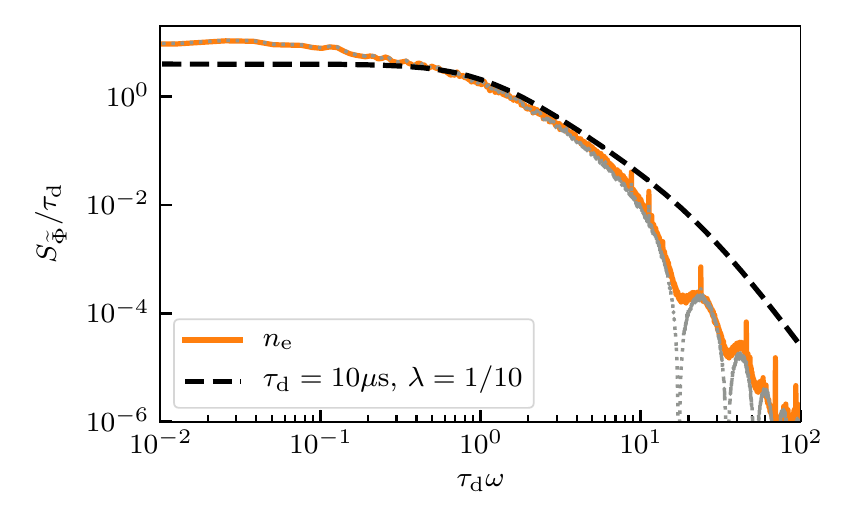}}
    \caption{Full line: Power spectral densities of MLP time series. Dotted line: Reconstructed
             time series from the RL-deconvolution. The black dashed line gives the power spectral
             density predicted by \Eqnref{PSD} for the two-sided exponential pulse.}
    \label{fig:deconv-compare-psd}
\end{figure}

The discrepancy between the low-frequency prediction of \Eqnref{PSD} and the PSD of the MLP data is
resolved by the deconvolution procedure. In \Figref{deconv-compare-psd}, the power spectral
densities of the MLP data time series are presented together with the power spectral densities of
the reconstructed time series and the analytic prediction. The reconstructed time series give the
same behavior for low frequencies as the MLP data, showing that this discrepancy is explainable by
the synthetic time series.

\begin{table}
    \begin{center}
    \begin{tabular}{c|c|c|c|c|c|c|}
        parameter & method & $I_{\mathrm{s}}$    & $n_{\mathrm{e}}$    & $T_{\mathrm{e}}$   & GPI $90\, cm$ & GPI $91\, cm$\\ \hline
$\gamma$ & PDF fit   & {1.01} &{3.22} & {1.33} & {3.54} & {2.01}\\ 
$\langle \tau_\mathrm{w} \rangle / \mu \mathrm{s}$ & RL deconv & { 8.6} & { 3.3} & { 7.1} & { 4.7} & { 9.0}\\ 
$\tau_\mathrm{d} / \mu \mathrm{s}$  & PSD fit    & { 9.2} & { 9.7} & { 9.8} & {19.7} & {19.1}\\ 
 $\gamma$    & $\tau_\mathrm{{d}} / \langle \tau_\mathrm{{w}} \rangle$    & ${ 1.1}$  & ${2.96}$    & ${1.39}$    & ${4.20}$    & ${2.12}$    
    \end{tabular}
    \caption{Process and pulse parameters estimated using MLP and GPI data sampled in discharge 3.}
    \label{tab:parameter_est}
    \end{center}
\end{table}

Table \ref{tab:parameter_est} summarizes the parameter estimation. The first three rows list
the parameters estimated using the methods described above. For the $\Isat$ and $\Te$ data, we find
$\gamma \approx 1$. This describes a strongly intermittent time series with significant quiet time in
between pulses. For the $\nee$ time series we find $\gamma \approx 3.2$, comparable to the estimates
for the GPI data. The average waiting time between pulses is $\avg{\tauw} \approx 8\, \mus$. The
best estimate for $\avg{\tauw}$ from the $\nee$ time series is given by $\avg{\tauw} \approx 3.3\, \mus$, 
estimates from the GPI data are larger by a factor of $2-3$, depending on the radial position of the
view. The pulse duration time for the MLP data is $\taud \approx 10\, \mus$, smaller by a factor of
two than for the GPI data, probably for the reasons discussed above.

The bottom row lists the intermittency parameter calculated using the estimated pulse duration time
and average waiting time, $\gamma = \taud / \avg{\tauw}$. The deconvolution algorithm uses $\taud$ 
from the power spectrum as an input parameter and $\gamma$ from the PDF fit as a constraint. 
Therefore, the fact that $\taud / \avg{\tauw}$ is comparable to $\gamma$ estimated from the PDF fit
is a good consistency check.

\section{Conclusions and summary}
Fluctuations of the scrape-off layer plasma have been studied for a series of ohmically heated
discharges in Alcator C-Mod. It is found that the radial variations of the lowest order statistical
moments, calculated from MLP and GPI measurements, are quantitatively similar. Time series data
from both MLP and GPI diagnostics, feature intermittent, large-amplitude bursts. As shown in numerous
previous publications, the time series are well described as a superposition of uncorrelated pulses
with a two-sided exponential pulse shape and a pulse amplitude that closely follows an exponential
distribution. In this contribution we demonstrate that the parameters which describe the various
parameters of the  stochastic process agree across MLP and GPI diagnostics. In particular, the same
statistical properties apply to the ion saturation current, electron density and temperature, and
the line emission intensity.

Radial profiles of the relative fluctuation level, skewness, and excess kurtosis, as estimated from
both MLP and GPI data, are of similar magnitude and are monotonically increasing with distance from
the LCFS. This holds regardless of using $\Isat$, $\nee$ or $\Te$ from the MLP.
For the GPI data the time series feature an intermittency parameter $\gamma \approx 2-3$,
when estimated from a fit on the PDF. Estimating the intermittency parameter by a fit on the PDF of 
the different MLP data time series yields $\gamma \approx 3$ for the $\nee$ data and 
$\gamma \approx 1$ for both the $\Isat$ and $\Te$ data. 
Pulse duration times, estimated from fits on the time series frequency power spectral density,
are $\taud \approx 10\, \mus$ for all MLP data time series while we find $\taud \approx 20\, \mus$
for the GPI data time series. This deviation by a factor of 2 is likely due to the relatively large
in-focus spot size of the individual GPI views. Reconstructing the distributions of waiting times between 
consecutive pulses from a Richardson-Lucy deconvolution, yields average waiting times between
pulses of $\avg{ \tauw } \approx (3,7,9)\, \mus$ for the $(\nee, \Te, \Isat)$ data. Using GPI data
time series, we find $\avg{ \tauw } \approx 5$ and $10\, \mus$ for the views at $R=90.7$ and
$91.0\, \mathrm{cm}$ respectively. We note that the GPI view at $R=91.0\,\mathrm{cm}$ is close to
the limiter shadow. Finally, estimating the intermittency parameter as $\taud / \avg{ \tauw }$ from
the deconvolution of the time series gives almost the same values as estimating
$\gamma$ by a fit on the PDF. These findings show that the model parameters of the stochastic model,
$\gamma$, $\taud$ and $\avg{ \tauw }$, are indeed a good parameterization of the plasma
fluctuations, independent of the diagnostic used to measure them. Reconstructing the arrival times 
and amplitude of the individual pulses using Richardson-Lucy deconvolution is an invaluable tool for
obtaining the distribution of waiting times between consecutive pulses.

Our analysis also suggests that calculating a synthetic line emission signal using the instantaneous
plasma parameters reported by the MLP results in a signal with different fluctuation statistics than
the time series actually measured by the GPI. The synthetic data time series present intermittent
pulses, but with a different shape than observed by the GPI. The PDF of these signals furthermore 
are close to a normal distribution, with low moments of skewness, excess kurtosis and no elevated
tails. We hypothesize argued that ionization, where hot plasma filaments locally decrease the puffed
gas density, is the main cause of this phenomenon and therefore should be accounted for in such an
attempt to reproduce the emission from measurements of $\nee$ and $\Te$.


Having established $\gamma$, $\taud$, and $\avg{\tauw}$ as consistent estimators for fluctuations 
in the scrape-off layer, future work will focus on describing their variations with plasma parameters.

\section{Acknowledgements}
This work was supported with financial subvention from the Research Council of Norway under Grant 
No. 240510/F20 and the from the U.S. DoE under Cooperative Agreement No. DE-FC02-99ER54512 40 using
the Alcator C-Mod tokamak, a DoE Office of Science user facility. A.T. and O.E.G. were supported by
the UiT Aurora Centre for Nonlinear Dynamics and Complex Systems Modelling. R.K., O.~E.~G., and
A.~T.~ acknowledge the generous hospitality of the MIT Plasma Science and Fusion Center during a
research stay where part of this work was performed. J.L.T acknowledges the generous hospitality of
UiT The Arctic University of Norway and its scientific staff during his stay there.

\appendix

\section{Local and global fluctuations} 

The emission intensity, measured by GPI, is often parameterized as
\begin{align}\label{eq:emission}
I & = n_0 \times F \left( \nee,\Te \right),
\end{align}
where $n_0$ is a constant neutral background density. Thus, the differential of $I$ can be written
as
\begin{align}
    \frac{\mathrm{d}I}{I} & = \frac{\partial \ln f}{\partial \ln \nee} \frac{\mathrm{d}\nee}{\nee} + \frac{\partial \ln f}{\partial \ln \Te} \frac{\mathrm{d}\Te}{\Te}, \label{eq:emission_differential}
\end{align}
where we use the notation $\partial \ln f(x) / \partial \ln x = \left( x/f(x) \right) \partial f(x) / \partial x$. 
Assuming small fluctuation amplitudes, the differential of a function $u$ can be approximated as
\begin{align}
\frac{\mathrm{d} u}{u} \approx \frac{\triangle u}{u} = \frac{u - \avg{u}}{\avg{u} + \triangle u} \approx \frac{u - \avg{u}}{\avg{u}}.
\end{align}
Here, $\triangle u$ is a small, but non-infinitesimal change in $u$ and $\avg{u}$ denotes an
average. That is, the relative, infinitesimal change in a function $u$ is approximately the
deviation of $u$ to an average $\avg{u}$ relative to this average. This approximation gives the
local density and temperature exponents $\beta_n$ and $\alpha_T$:
\begin{equation}
    \frac{I-\avg{I}}{\avg{I}} \approx \beta_n \frac{\nee-\avg{\nee}}{\avg{\nee}} + \alpha_T \frac{\Te - \avg{\Te}}{\avg{\Te}},
\end{equation}
where $\beta_n = \partial \ln f / \partial \ln \nee$ and $\alpha_T = \partial \ln f / \partial \ln \Te$ at a given (fixed) $\avg{\nee}$ and $\avg{\Te}$.

For large deviations relative to the mean values, this local approximation breaks down for two
reasons. First, the infinitesimal change $\text{d} u$ can no longer be approximated as a variation
relative to a mean value. Second, the partial derivatives in \Eqnref{emission_differential}, which
are evaluated at a fixed point, are not necessarily constant when using non-infinitesimal values for
the $\mathrm{d} \nee$ or $\mathrm{d} \Te$. The local exponents are therefore not constant, and the
full, global \Eqnref{emission} must be used.

\bibliographystyle{jpp}

\bibliography{myrefs.bib,deconvrefs.bib}

\end{document}